\documentclass[aps, 10pt, prd,notitlepage,twocolumn, superscriptaddress]{revtex4-2}

\usepackage{amsmath,amssymb,amsfonts}
\usepackage[utf8]{inputenc}
\usepackage[T1]{fontenc}
\usepackage{mathrsfs}
\usepackage[colorlinks=true]{hyperref}
\hypersetup{citecolor=cyan,linkcolor=magenta}
\usepackage{anyfontsize}
\usepackage[usenames,dvipsnames]{xcolor}

\usepackage{newtxtext}
\usepackage{newtxmath}
\usepackage{tensor}
\usepackage{bm}

\usepackage{graphicx}

\usepackage{natbib}

\usepackage[normalem]{ulem}

\usepackage{journals}

\allowdisplaybreaks[1]

\begin{document}

\title{Gravitational waves from accretion-induced descalarization in massive scalar-tensor theory}

\author{Hao-Jui Kuan}
\email{hao-jui.kuan@uni-tuebingen.de}
\affiliation{Theoretical Astrophysics, Eberhard Karls University of T\"ubingen, T\"ubingen 72076, Germany}
\affiliation{Department of Physics, National Tsing Hua University, Hsinchu 300, Taiwan}

\author{Arthur G. Suvorov}
\affiliation{Manly Astrophysics, 15/41-42 East Esplanade, Manly, NSW 2095, Australia}
\affiliation{Theoretical Astrophysics, Eberhard Karls University of T\"ubingen, T\"ubingen 72076, Germany}

\author{Daniela D. Doneva}
\affiliation{Theoretical Astrophysics, Eberhard Karls University of T\"ubingen, T\"ubingen 72076, Germany}
\affiliation{INRNE - Bulgarian Academy of Sciences, 1784  Sofia, Bulgaria}

\author{Stoytcho S. Yazadjiev}
\affiliation{Theoretical Astrophysics, Eberhard Karls University of T\"ubingen, T\"ubingen 72076, Germany}
\affiliation{Department of Theoretical Physics, Faculty of Physics, Sofia University, Sofia 1164, Bulgaria}
\affiliation{Institute of Mathematics and Informatics, 	Bulgarian Academy of Sciences, 	Acad. G. Bonchev St. 8, Sofia 1113, Bulgaria}

\date{\today}

\begin{abstract}
\noindent Many classes of extended scalar-tensor theories predict that dynamical instabilities can take place at high energies, leading to the formation of scalarized neutron stars. Depending on the theory parameters, stars in a scalarized state can form a solution-space branch that shares a lot of similarities with the so-called mass twins in general relativity appearing for equations of state containing first-order phase transitions. Members of this scalarized branch have a lower maximum mass and central energy density compared to Einstein ones. In such cases, a scalarized star could potentially over-accrete beyond the critical mass limit, thus triggering a gravitational phase transition where the star sheds its scalar hair and migrates over to its non-scalarized counterpart. Such an event resembles, though is distinct from, a nuclear or thermodynamic phase transition. We dynamically track a gravitational transition by first constructing hydrostatic, scalarized equilibria for realistic equations of state, and then allowing additional material to fall onto the stellar surface. The resulting bursts of monopolar radiation are dispersively-stretched to form a quasi-continuous signal that persists for decades, carrying strains of order $\gtrsim10^{-22} (\text{kpc}/L)^{3/2} \text{ Hz}^{-1/2}$ at frequencies of $\lesssim300 \text{ Hz}$, detectable with the existing interferometer network out to distances of $L\lesssim10 \text{ kpc}$, and out to a few hundred kpc with the inclusion of the Einstein Telescope. Electromagnetic signatures of such events, involving gamma-ray and neutrino bursts, are also considered.
\end{abstract}


\maketitle

\textit{Introduction.}
General relativity (GR) has historically provided an excellent description for both local (e.g., solar system) and global (e.g., cosmology) gravitational phenomena. It is well known however that the theory cannot by itself be fully complete, and the non-renormalizability of the action implies that additional ingredients, possibly in the form of non-minimally coupled fields \cite{brav85,ruf18}, should activate at extreme scales. Nevertheless, any theoretical extension must be virtually invisible at low energies, and also somehow suppressed in certain strong-field environments. For example, binary pulsar experiments restrict the possibility for significant sub-quadrupolar radiation over super-Compton length scales \cite{fre12,als15,kram21,Shao17,Zhao22}, and gravitational-wave (GW) experiments suggest that (at least some) black holes should be approximately, if not exactly, Kerr \cite{ligo15}. An observationally-viable class of extensions that can survive these issues is massive scalar-tensor theory (STT): the mass of the scalar field suppresses the scalar dipole radiation \cite{Ramazanoglu16,Yazadjiev16}, and the classical no-hair theorems tend to be respected \cite{herd15}, implying that astrophysically stable black holes would be indistinguishable from their GR counterparts.

Material degrees of freedom in these theories however allow for the possibility of \emph{scalarized} stars \cite{dam93}. This phenomenon can be generally thought of as a consequence of the effective curvature-coupled mass term, which appears in the relevant Klein-Gordon equation, changing sign once some critical threshold is breached, thereby inducing a tachyonic instability (though see also Ref. \cite{Doneva:2021tvn}). 
Thus for neutron stars a critical compactness exists at which a branch of strongly-scalarized solutions emerges. In some cases, the heaviest scalarized neutron star has lower baryon mass and central energy density compared to the maximum mass (stable) non-scalarized neutron star \cite{Sotani:2017pfj} and there is a gap between the two branches where no stable neutron star solutions exist. This picture resembles very closely the so-called mass twins in pure GR that are manifestations of the presence of a first order phase transition in the equation of state (EOS) \cite{kam81,Glendenning:1998ag,sch00,shaf02}. This implies that if a near-critical scalarized star were to acquire additional mass through accretion, the system may promptly discharge its scalar hair. Importantly, the neutron star does not need to collapse to a black hole in this scenario, as considered in, e.g., \cite{Novak:1999jg,Novak:1997hw,rot20,Mendes21}, but rather may undergo a \emph{gravitational} phase transition, distinct from a \emph{material} (e.g., hadron-quark) phase transition \cite{chat20}, and migrate to the GR branch pertaining to the same EOS. This novel scenario is considered in this Letter.

A migration to a new branch is likely to carry a variety of observational signatures. As a scalar shedding necessarily compactifies the star over a short, dynamical timescale ($\gtrsim$~ms), abrupt changes in the electromagnetic output of the source, most notably associated with gamma-ray burst (GRB) afterglows and neutrino bursts \cite{zhang01,bern20,sar20}, would point towards such a transition. These signatures could, however, be imitated by a nuclear phase transition \cite{Most:2018eaw,Bauswein:2018bma,Weih:2019xvw,Liebling:2020dhf,Most:2019onn,Blacker:2020nlq}, thus highlighting the well-known degeneracy between effects coming from a modification of gravity and the uncertainties in the nuclear EOS (e.g., \cite{Berti:2015itd,shao19}). One key difference is that a gravitational transition will unleash a burst of scalar radiation which, for a massive scalar field theory, will be dispersively stretched into a quasi-continuous signal, as higher frequency components are first to arrive at the detector(s) \cite{rot20}. GW afterglows lasting up to a $\sim$kyr may therefore follow a gravitational phase transition.

\textit{Formalism and equations of motion.} The action of a STT of whichever flavor [e.g., Brans-Dicke, Bergmann-Wagoner, or even $f(R)$ theories] can be transformed into the Einstein frame
\begin{equation}
\begin{aligned}
\hspace{-0.12cm}	S=\int\frac{\sqrt{-g}}{16\pi}d^4x \Big(R-2\partial_{\mu}\varphi\partial^{\mu}\varphi - 4V \Big) + S_{M}[\Psi,A^2g_{\mu\nu}],
\end{aligned}
\end{equation}
for matter portion $S_{M}$, metric tensor $g$, scalar field $\varphi$, Ricci scalar $R$. Scalar field potential is taken as $V(\varphi)=m_{\varphi}^2\varphi^2/2$ \cite{Sperhake:2017itk}, whose saddle point poses the boundary condition $\varphi(r\rightarrow\infty)=0$ for $\varphi$. The transition to the physical Jordan frame metric $\tilde{g}_{\mu\nu}$ is done via a Weyl scaling $\tilde{g}_{\mu\nu}=A(\varphi)^2g_{\mu\nu}$, where we adopt the spherically-symmetric Jordan frame metric, $\tilde{g}_{\mu\nu}=\text{diag} [-\alpha^{2},X^{2},A(\varphi)^2r^{2},A(\varphi)^2r^{2}\sin^{2}\theta ]$, and choose the conformal factor $A(\varphi)=\exp(\alpha_0\varphi+\beta_0\varphi^2/2)$ following \cite{Gerosa16,Sperhake:2017itk}. The resulting field equations are given in Appendix~\ref{appendix.A}, while we note here that the term responsible for scalarization is proportional to the logarithmic derivative of the conformal factor $\alpha(\varphi)=d\ln A/d\varphi$. The choice we make, $\alpha(\varphi)=\alpha_0+\beta_0 \varphi$, therefore corresponds to the two leading terms in the Maclaurin expansion of any regular function; higher-order terms do not change the picture of scalarization qualitatively for a large class of more complicated conformal factors because $\varphi \ll 1$ everywhere \cite{dam93,alta17,Chiba22}. Simulations for another coupling function are shown in Appendix~\ref{appendix.E} to demonstrate that the phenomenon put forward in this Letter is not quenched by higher-order effects.

Two additional GW modes beyond those in GR are raised by $\varphi$, viz.~the breathing (`B') and longitude (`$\ell$') modes, which carry the same response functions up to a sign flip (see ~Eq.~(134) of \cite{Will14}). The strain of the latter is weaker by a factor $(\lambda_{\varphi}f)^{-2}$ relative to the former, which reads $h_{\varphi} = 2\alpha_0 \varphi$. The strain felt by a LIGO-Virgo-like array (two orthogonal antennas), at a distance $L$ from the source, is thus
\begin{align}\label{eq:strain}
    h(L,t) = h_{\varphi}(L,t)\{1-[\lambda_{\varphi}f(L,t)]^{-2}\}/2,
\end{align}
when the source orients optimally, where $\lambda_{\varphi}=2\pi\hbar/m_{\varphi}$ is the Compton length-scale for the massive scalar field, $t$ is the retarded time post-emission, and $f(L,t)$ is the characteristic frequency of the signal. Furthermore, modes with distinct frequencies propagate at different, subluminal velocities, and the full power-spectral density (PSD), $2\sqrt{f}|\tilde{h}(f)|$, will not arrive at the detector simultaneously. As a result, the dispersively-stretched burst becomes quasi-monochromatic over a $L$- and $m_{\varphi}$-dependent timescale \cite{Sperhake:2017itk,rot20}. Therefore, there is an implicit time-dependence, encoded in $f$, for the witness. The quasi-monochromatic feature implies that a phase-coherent search can be implemented, and the signal-to-noise ratio (SNR), $4\int df\big[|\tilde{h}(f)|^2/S_n(f)\big]$, can be obtained by integrating the strain over a narrow frequency interval. In the limit $Tf\gg1$, i.e., when many cycles are observed, the SNR can be obtained by dividing the \emph{effective} PSD from the aforementioned time-domain integration \cite{rot20},
\begin{equation}\label{eq:SNR}
    \sqrt{S_f}=\sqrt{T}\alpha_0A(L,t) \{1-[\lambda_{\varphi}f(L,t)]^{-2}\},
\end{equation}
by the noise spectral curve $\sqrt{S_n(f)}$, where we assume an optimally-oriented detector. Here the amplitude is $A(L,t)\simeq\mathcal{A}(f)L^{-3/2}\lambda_\varphi\left(f^2-\lambda_\varphi^{-2}\right)^{3/4}$ with $\mathcal{A}(f)$ the Fourier component of the scalar-GW extracted at some distance $\lambda_{\varphi}<r_{\text{out}}\ll L$ so that it contains the wave content that eventually propagates to a detector at $L$ (cf.~Eq.~(57) in \cite{rot20}, while noting that our definition for $A$ differs from theirs by a factor of $L^{-1}$).

\begin{figure}
	\includegraphics[width=\columnwidth]{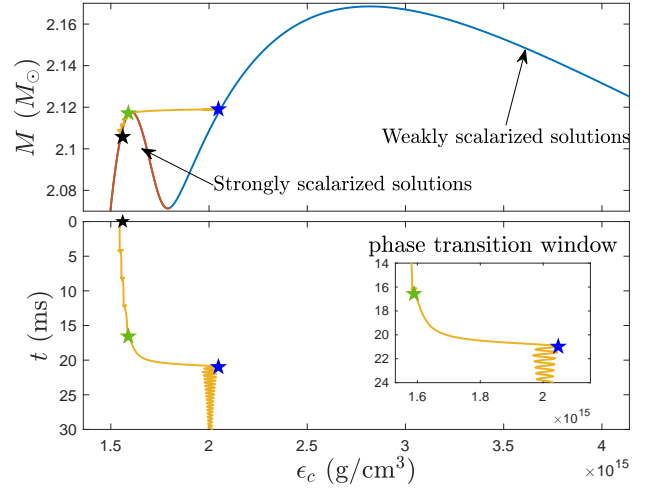}
	\caption{Evolutionary track of an APR4, near-critical scalarized star under accretion: gravitational mass, $M$, as a function of central energy density, $\epsilon_c$ (top panel), together with the time evolution of $\epsilon_{c}$ itself (bottom panel). The blue branch represents weakly-scalarized stars, while the red branch is strongly-scalarized.
	The green and blue stars mark the onset and the termination of descalarization, respectively, while the black star marks the initial state of the accretion simulation described in the main text. 
	}
	\label{fig:track}
\end{figure}

\textit{Scalarized neutron stars.} 
In the present Letter, we adopt a piecewise-polytropic approximation to the APR4 EOS \cite{OBoyle20}, which withstands constraints coming from GW 170817 \cite{LVK} and supports masses which accommodate the heaviest neutron star observed to-date, viz.~PSR J0740+6620 $(M = 2.14^{+0.10}_{-0.09} M_{\odot})$ \cite{NANOGrav}. 
Specifics related to this EOS and its numerical implementation are given in Appendix~\ref{appendix.B}.
In the considered STT and for large enough stellar compactness, the neutron star scalarizes, meaning it develops a strong, localized scalar field. If $\alpha_{0} \neq 0$, purely GR solutions do not exist within the theory, and all stars have at least some tiny residual $\varphi \neq 0$, i.e., they must be at least weakly scalarized. For practical purposes, however, weakly-scalarized solutions are virtually indistinguishable from GR counterparts at high densities and thus with a slight abuse of language, we call these solutions \emph{descalarized}.

A sequence of neutron star solutions is shown in the top panel of Fig.~\ref{fig:track} as a function of central energy density $\epsilon_c$ for $\alpha_0=10^{-2}$, $\beta_0=-5$, and $m_\varphi=10^{-14}$ eV. This value of $m_\varphi$ mitigates the tension with binary-pulsar constraints, since radiation is suppressed over super-Compton length-scales $r\gg\lambda_{\varphi}$ \cite{als15,Yazadjiev16,Ramazanoglu16}, and effectively allows for a broad range of $\beta_0$ \cite{Tuna:2022qqr}. The other parameters, which similarly respect binary pulsar constraints, are chosen such that the maximal mass for the scalarized branch ($2.118M_{\odot}$; red curve) is less than that of the `weakly-scalarized' branch ($2.168M_{\odot}$; blue curve). Moreover, for both the red and the blue branches the neutron star solutions are stable up to the maximum mass point and lose stability afterwards \cite{Harada98,Mendes:2018qwo,Mendes21}, resulting in a picture reminiscent of the so-called mass twins in pure GR \cite{kam81,sch00,Glendenning:1998ag,shaf02}. Thus, a phase transition (descalarization) from the red to the blue branch can be realized if additional mass is added. For the APR4 EOS, this scenario is possible provided that $-5.41 \lesssim \beta_{0} \lesssim -4.78$, with the exact range depending on $\alpha_{0}$ and the coupling functions $V(\varphi)$ and $A(\varphi)$. Similar ranges apply for other EOS. These limits will differ also if we consider scalarization in more general theories of gravity, such as tensor-multi-scalar theories \cite{Damour92,Horbatsch:2015bua,Doneva:2020afj}.

\textit{Accretion dynamics.}
A neutron star that over-accretes beyond the peak of the scalarized curve, displayed by the green star in Fig.~\ref{fig:track}, will undergo a phase transition by descalarizing. This scenario may occur either for a newborn star after a merger or collapse through fallback accretion, or a mature star in a binary undergoing Roche-lobe overflow. In the former case, debris disks containing $\lesssim 0.2 M_{\odot}$ worth of material \cite{bern20}, though potentially much more in a core-collapse \cite{piro11}, will form around the birth site. A sizeable fraction $(\lesssim 40\%)$ of it may eventually fall back onto the stellar surface \cite{ish21}. Accreted masses may total $\lesssim0.8M_{\odot}$ in some X-ray binaries \cite{van95}, though such amounts can accumulate only over long (potentially $\sim$ Gyr) timescales. The details of the accretion process itself are complicated however, since the neutron star may be spinning rapidly enough that material is repelled by a centrifugal barrier (`propeller' effect \cite{piro11}), pressure gradients from nucleosynthetic heating can accelerate ejecta before it has a chance to return \cite{des19}, and material will not fall isotropically onto the surface but rather may be guided onto the magnetic poles by the Lorentz force \cite{lamb73}.

In this Letter however, our main goal is not to simulate a realistic accretion process in a STT, but rather to illustrate qualitatively how the dynamical acquisition of additional mass can trigger a descalarization. To this end, accretion is artificially simulated by superposing a radial, Gaussian bulk centred at $0.9R_{\star}$, for stellar radius $R_{\star}$, with a width (``standard deviation'') of $1$ km, every $4$ ms. The process is then halted when a total (baryon) mass of $0.015M_{\odot}$ has been added (after $16.01$ ms). The average accretion rate of $\simeq 0.94M_{\odot}\text{s}^{-1}$ is marginally slower than that observed in the first few ms of merger simulations, viz. $\lesssim 1M_{\odot}\text{s}^{-1}$ (Fig.~7 of \cite{Fujibayashi17}). Using this scheme, we model a dynamical descalarization. As shown in Fig. \ref{fig:track}, the system begins in a particular state (shown by the black star), gains some mass (green star), and the descalarization process begins (orange line), until eventually the system oscillates around a certain, stable state on the new branch (blue star). The bottom panel of Fig.~\ref{fig:track} shows the evolution of $\epsilon_c$ in this example; the increase of $\epsilon_{c}$ from $1.6 \times 10^{15}\, \mbox{g cm}^{-3}$ to $\sim 2 \times 10^{15}\, \mbox{g cm}^{-3}$ in $\sim 4\, \mbox{ms}$ indicates that a rapid compactification accompanies descalarization.

We have verified that the (post-)descalarization dynamics remain the same if the bulk is accreted with longer waiting time, and that our results are not overly sensitive to the particulars of the chosen accretion profile, as described above, by studying other bulk impositions (see Appendix~\ref{appendix.D}). Nevertheless, we stress that our profiles are not representative of realistic astrophysical processes, though they allow us to capture the salient features of a gravitational phase transition. Importantly, magnetic fields only couple weakly to the scalar sector, and thus even if the geometry of the accreted-mass buildup (`mountain') is sensitive to the former (e.g., \cite{suv21}), the descalarization dynamics, and resulting scalar-GW signal, are not. The latter aspects are discussed below.

\textit{Results.} 
Introducing the auxiliary variables $\psi=\alpha^{-1}(\partial\varphi/\partial t)$ and $\eta=X^{-1}(\partial\varphi/\partial r)$,
we have that the energy $E_{\varphi}$ and luminosity $\mathcal{L}_{\varphi}$ of the scalar field read (see Appendix~\ref{appendix.C})
\begin{eqnarray}
	E_{\varphi} = \int dr \left[ \frac{r^2A(\varphi)^2}{2}(\psi^{2}+\eta^{2})+V \right], 
	\label{eq:s_eng}
\end{eqnarray}	
and
\begin{eqnarray}
    \mathcal{L}_{\varphi} = A(\varphi)^{-2}r^2X\alpha\psi\eta, 
    \label{eq:s_lum}
\end{eqnarray}	
which defines the corresponding energy leakage as
\begin{align}\label{eq:s_gw}
 	E^{\text{(scalar)}}_{\text{GW}} = \int \mathcal{L}_{\varphi} dt.
\end{align}
For the simulation shown in the top panel of Fig.~\ref{fig:track}, the scalar energy \eqref{eq:s_eng} plummets to zero, from its initial value of $0.051M_{\odot}$, after descalarization. The near zone $(r\ll\lambda_\varphi)$ extraction of $E_{\text{GW}}^{(\text{scalar})}$ suggests an energy loss $\gtrsim40$ times less than the decrease of $E_{\varphi}$, indicating that most of the scalar energy transforms into gravitational binding energy since the stellar radius shrinks from $11.56$ km to $10.36$ km between the initial and final states, while the (gravitational) mass increases by $\approx 0.013M_{\odot}$. After the emission propagates to distances comparable to $\lambda_\varphi$, the dispersion suppresses the low frequency component(s) and leads to a stretching of the waveform. The scalar energy leakage ultimately saturates at super-Compton length-scales, where a reduction of a factor of $\sim 2$ in the near-zone radiation is seen due to the dispersion.

\begin{figure}
	\includegraphics[width=\columnwidth]{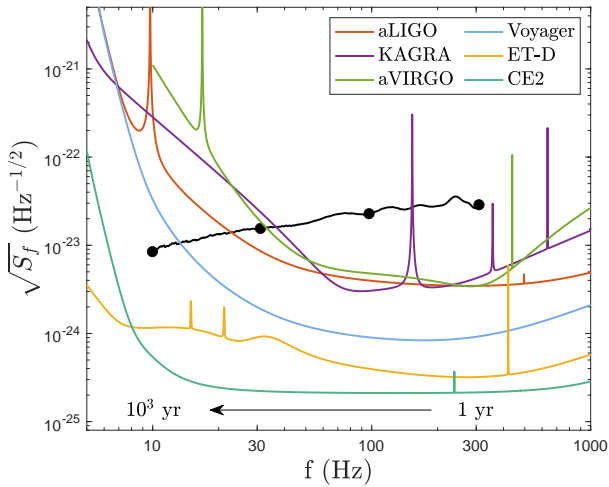}
	\caption{
	Effective, root PSD $\sqrt{S_f}$ [Eq.~\eqref{eq:SNR}] at $L=10$ kpc and $T$ = 2 months for the strain of the scalar-induced GW mode as a function of retarded time from 1 to $10^3$ years (black curve; the $k$-th dot from the right along the curve represents $10^{(k-1)}$ years). Overlaid are the sensitivity curves of existing and upcoming GW interferometers.}
	\label{fig:miscellaneous}
\end{figure}

Assuming an observation duration of $T=2$ months, over which the signal evolves slowly for $m_{\varphi}=10^{-14}$ eV \cite{rot20}, we present the numerical evaluation of the signal amplitude \eqref{eq:SNR} for scalar GWs for $\alpha_{0} = 10^{-2}$ at a fixed distance $L = 10\, \mbox{kpc}$ in Fig. \ref{fig:miscellaneous}. 
The $k$-th notch, starting from the right, plotted over the signal (black curve) stands for $t=10^{(k-1)}$ years, illustrating that the bulk of the signal persists for $\lesssim$ centuries. We see that for $T =2$ months and these theory parameters (see also below), the signal should be detectable with sufficiently high signal-to-noise ratio with the existing interferometer network out to distances of $L\lesssim10 \text{ kpc}$, and out to a few hundred kpc with the inclusion of the Einstein Telescope (ET). Note that the distance and observation time $T$ are, to a large degree, degenerate. We find that increasing $T$ by a factor $\sim 2$ allows for events at distances a factor $\sim 2^{1/3}$ further out to become visible, assuming a narrow-band search is carried out.

To quantify the detectability in general, we compared outputs from a variety of simulations with varying $\alpha_{0}$. 
We find a fitting to the root of the \emph{effective} PSD, in units of $\text{Hz}^{-1/2}$, as
\begin{equation} \label{eq:estim}
\hspace{-0.25cm}\frac{\sqrt{S_f}}{10^{-23}} \approx 3 
\left(\frac{\alpha_0}{10^{-2}}\right) 
\left(\frac{T}{5\times10^6\text{ s}}\right)^{\frac{1}{2}} 
\left(\frac{10\text{kpc}}{L}\right)^{\frac{3}{2}} 
\left(\frac{f}{300\text{Hz}}\right)^{0.34},
\end{equation}
with the frequency of the signal approximated as $f(L,t)\approx2.42(t+L)/\sqrt{t(t+2L)}$ Hz (cf.~Eq.~(53) in \cite{rot20}). We note that larger $T$ may be used for greater retarded times since the timescale for the frequency evolution, $f/\dot{f}$, scales as $t$, thus opening the possibility for much larger \emph{effective} PSD. In particular, since the signal \eqref{eq:estim} is quasi-continuous, an extended narrow-band search could be carried out if one knew when the system descalarized, as the dispersion relation directly equates the relative delay with a frequency. Furthermore, multiple sensors can act to `fuse' data together in a way that improves the overall signal-to-noise ratio beyond that inferred from equation \eqref{eq:estim} (see Sec.~IV.~E.~of \cite{szcz21} for a detailed comparison of achievable sensitivities with different networks). 

\textit{Connection to matter phase transitions.}
The phase transitions from scalarized to non-scalarized states considered here bear striking similarities to the material phase transitions from confined hadronic to deconfined quark matter. In both cases, there can be stars of equal mass but different radii, that are separated by a range of central energy densities where the stable solution space is empty (i.e., twin stars \cite{kam81,Glendenning:1998ag,sch00,shaf02}). The astrophysical implications of matter phase transitions, and especially the GW signatures, have attracted considerable attention recently (see, e.g., \cite{Most:2018eaw,Bauswein:2018bma,Weih:2019xvw,Liebling:2020dhf,Most:2019onn,Blacker:2020nlq}). In each case there will be a descalarization analogue, with the main difference being an additional channel for energy loss -- the scalar radiation. 

As a proof of principle we concentrate on accretion in this Letter, however such analogues can be found also in cases without accretion. For a hot, newborn neutron star with an EOS that permits negatively charged, non-leptonic particles (e.g., hyperons or quarks), the hydrostatic support available to the star will reduce when neutrinos diffuse out of the core \cite{prak06}. This can lead to a delayed phase transition with a number of interesting observational signatures \cite{Weih:2019xvw}. 
A descalarization analogue of this delayed transition exists: depending on the chemical composition and theory parameters, the scalarized star may migrate to a non- or weakly-scalarized branch when the temperature drops below a critical threshold. 
A similar picture exists for cases where the star is centrifugally or magnetically supported: spindown or field decay reduces the maximum mass of the system, which could force the star to transition \citep{glen97,sg21}. Studying these processes in detail lies beyond the scope of the present paper, though complementing scalar-flavour phase transitions with studies of neutron star mergers in STT is likely to offer rich phenomenology as concerns the evolutionary track of neutron stars. This will be on one hand due to the additional channel of energy loss that, even if not detectable, will alter the merger remnant evolution. Some properties of the post-merger remnant, such as its oscillations frequencies, can also differ from GR due to the scalarization-related changes in stellar structure \cite{Shibata:2013pra}.

We point out that we discuss twin stars only as an interesting analogy with the observed process of descalarization. Our simulations and the predicted observational signatures are completely independent of the existence of such stars (the astrophysical relevance of twin stars is discussed in, e.g., \cite{Christian:2020xwz,Deloudis:2021agp,Espino:2021adh,Bauswein:2022vtq}).

\textit{Discussion and observational prospects.} While a detection of scalar GWs of the form shown in Fig. \ref{fig:miscellaneous} could be used to unambiguously identify that a descalarization took place, (massive) STTs may already leave traces in the events that lead up to the transition. A promising avenue for the formation of scalarized stars, which are also prone to over-accretion and descalarization, comes from binary mergers. The scalar field associated with the binary constituents may become excited during inspiral, leaving a clear imprint on the GW signal by accelerating the coalescence \cite{Shibata:2013pra} (see also \cite{bar13,pon14,Sagunski:2017nzb,huang19,huang21}). Despite progress though, certain key effects, such as rotation (though see \cite{oko20,oko20b}), are still missing from numerical simulations of mergers involving stars in STTs. This means that direct waveform comparisons with observed inspirals cannot be achieved yet. On the electromagnetic side, however, binary neutron-star merger events are also the progenitors for short GRBs, which offer avenues for indirectly observing a descalarization.

Many GRBs exhibit extended emissions at short-wavelengths following the main burst. Emission profiles that display a long-lived X-ray `plateau' are suggestive of persistent energy injections (`magnetar wind') from a massive, newborn neutron star \cite{zhang01,bern20}. Suppose that tensorial GWs were coincidentally observed with a short GRB (as occurred for GW170817 \cite{abb17}), followed by a plateau-like X-ray afterglow. The detection of a \emph{scalar} GW afterglow some time after the main event, which may persist for $\lesssim$ centuries (see Fig. \ref{fig:miscellaneous}), would clearly indicate that the remnant peeled its scalar hair. Even without such measurements, the nature of the electromagnetic afterglow will be affected by a scalar shedding as the star condenses (see Fig. \ref{fig:track}). The spindown power associated with magnetic dipole braking scales as $L_{\text{dip}} \propto R_{\star}^{6}$ (e.g., \cite{bern20}), and so a decrease in $R_{\star}$ by $\sim 5\%$ may then lead to a drop in the X-ray flux by $\lesssim 30\%$ over the descalarization timescale $(\sim 5 \text{ ms})$. Afterglow light-curves in this case may appear as `broken plateaus', like that of GRB 170714A \cite{hou18}. Conservation of angular momentum however implies that the star should spin-up as a result of descalarization, and thus the drop may be less pronounced because $L_{\text{dip}} \propto \Omega^4$. Likewise, the temperature of the star should increase from the compactification. Magnetohydrodynamic processes involving magnetic field reorganization may also take place, extending the dip timescale and enriching the phenomenology.

A descalarization-induced compactification may itself instigate a nuclear phase transition (e.g., quark deconfinement) due to the sudden increase in the core density \cite{prak06,Weih:2019xvw}. Alternatively, the scalarized neutron star will collapse to a hairless black hole if no stable branch is available. From a scalar-GW perspective, these events would be indistinguishable \cite{Mendes21}, though could be told apart via the nature of the X-ray afterglow. If emissions persisted after the scalar energy release, a gravitational phase-transition would be the favoured scenario since black hole formation, which effectively terminates the stellar wind that is pumping radiation energy into the forward shock, should instead manifest as a sharp drop in the flux (as is often observed \cite{sar20}).

The closest GRBs that have thus far been observed are GRBs 980425 and 170817A at distances of $\sim 40$ Mpc \cite{gal98,abb17}. This distance is a factor $\sim 4000$ times larger than that plotted in Fig.~\ref{fig:miscellaneous}. As such, unless $\alpha_{0}$ is $\gtrsim 10$ times bigger than the value we have used and year-long ($T \gtrsim \text{yr}$) searches are carried out, we are unlikely to observe this scenario in its full capacity even with Cosmic Explorer (CE) \cite{Reitze:2019iox} or ET \cite{Punturo:2010zz,Hild:2010id,Maggiore:2019uih} because of the $L^{-3/2}$ dependence in the effective PSD \eqref{eq:estim}, should such stars exist. Other multi-messenger possibilities for identifying a neutron star post-descalarization come from neutrino bursts (from Urca cooling or shocks triggered by compactification; cf. \cite{sag09}) or indeed a burst of GWs (if the now descalarized star collapses) at some later time, either of which would again be hard to explain with a black hole remnant. It is also not necessarily the case that a neutron star must descalarize shortly after birth. Mature stars residing in the disks of active galactic nuclei or high-mass X-ray binaries \cite{qin98} are particularly disposed to over-accretion. Accretion-induced collapse rates could reach $\lesssim 20 \text{ Gpc}^{-3} \text{ yr}^{-1}$ from the former channel \cite{perna21}. Descalarizations of Galactic stars via the latter channel should be observable with high SNR by ET. Overall however, in the absence of a detection of scalar GWs, one may not be able to tell whether a phase transition was of a nuclear or gravitational nature. This exemplifies further the well-known degeneracy between modifications of gravity and EOS uncertainty \cite{Shibata:2013pra,Yazadjiev16,shao19}.

\section*{Acknowledgements}
We are grateful to J{\"u}rgen Schaffner-Bielich for carefully reading the manuscript and providing valuable advice which improved the quality of this Letter. We would also like to thank Jan-Erik Christian and David Blaschke for helpful discussion concerning twin stars. The anonymous referees are appreciated for providing helpful feedback.
AGS recognises funding received from the European Union's Horizon 2020 Programme under the AHEAD2020 project (grant n. 871158). DD acknowledges financial support via an Emmy Noether Research Group funded by the German Research Foundation (DFG) under grant no. DO 1771/1-1. SY would like to thank the University of T{\"u}bingen for the financial support. The partial support by the Bulgarian NSF Grant KP-06-H28/7 is acknowledged. 


\bibliographystyle{unsrt}
\bibliography{references}

\begin{thebibliography}{10}

\bibitem{brav85}
A.~O. Barvinsky and G.~A. Vilkovisky.
\newblock {The Generalized Schwinger-Dewitt Technique in Gauge Theories and
  Quantum Gravity}.
\newblock {\em Phys. Rept.}, 119:1--74, 1985.

\bibitem{ruf18}
Michael~S. Ruf and Christian~F. Steinwachs.
\newblock {One-loop divergences for $f(R)$ gravity}.
\newblock {\em Phys. Rev. D}, 97(4):044049, 2018.

\bibitem{fre12}
Paulo C.~C. Freire, Norbert Wex, Gilles Esposito-Farese, Joris P.~W. Verbiest,
  Matthew Bailes, Bryan~A. Jacoby, Michael Kramer, Ingrid~H. Stairs, John
  Antoniadis, and Gemma~H. Janssen.
\newblock {The relativistic pulsar-white dwarf binary PSR J1738+0333 II. The
  most stringent test of scalar-tensor gravity}.
\newblock {\em Mon. Not. Roy. Astron. Soc.}, 423:3328, 2012.

\bibitem{als15}
Justin Alsing, Emanuele Berti, Clifford~M. Will, and Helmut Zaglauer.
\newblock {Gravitational radiation from compact binary systems in the massive
  Brans-Dicke theory of gravity}.
\newblock {\em Phys. Rev. D}, 85:064041, 2012.

\bibitem{kram21}
M.~Kramer et~al.
\newblock {Strong-Field Gravity Tests with the Double Pulsar}.
\newblock {\em Phys. Rev. X}, 11(4):041050, 2021.

\bibitem{Shao17}
Lijing Shao, Noah Sennett, Alessandra Buonanno, Michael Kramer, and Norbert
  Wex.
\newblock {Constraining nonperturbative strong-field effects in scalar-tensor
  gravity by combining pulsar timing and laser-interferometer
  gravitational-wave detectors}.
\newblock {\em Phys. Rev. X}, 7(4):041025, 2017.

\bibitem{Zhao22}
Junjie Zhao, Paulo C.~C. Freire, Michael Kramer, Lijing Shao, and Norbert Wex.
\newblock {Closing the spontaneous-scalarization window with binary pulsars}.
\newblock 1 2022.

\bibitem{ligo15}
B.~P. Abbott et~al.
\newblock {Tests of general relativity with GW150914}.
\newblock {\em Phys. Rev. Lett.}, 116(22):221101, 2016.
\newblock [Erratum: Phys.Rev.Lett. 121, 129902 (2018)].

\bibitem{Ramazanoglu16}
Fethi~M. Ramazano\u{g}lu and Frans Pretorius.
\newblock {Spontaneous Scalarization with Massive Fields}.
\newblock {\em Phys. Rev. D}, 93(6):064005, 2016.

\bibitem{Yazadjiev16}
Stoytcho~S. Yazadjiev, Daniela~D. Doneva, and Dimitar Popchev.
\newblock {Slowly rotating neutron stars in scalar-tensor theories with a
  massive scalar field}.
\newblock {\em Phys. Rev. D}, 93(8):084038, 2016.

\bibitem{herd15}
Carlos A.~R. Herdeiro and Eugen Radu.
\newblock {Asymptotically flat black holes with scalar hair: a review}.
\newblock {\em Int. J. Mod. Phys. D}, 24(09):1542014, 2015.

\bibitem{dam93}
Thibault Damour and Gilles Esposito-Farese.
\newblock {Nonperturbative strong field effects in tensor - scalar theories of
  gravitation}.
\newblock {\em Phys. Rev. Lett.}, 70:2220--2223, 1993.

\bibitem{Doneva:2021tvn}
Daniela~D. Doneva and Stoytcho~S. Yazadjiev.
\newblock {Beyond the spontaneous scalarization: New fully nonlinear mechanism
  for the formation of scalarized black holes and its dynamical development}.
\newblock {\em Phys. Rev. D}, 105(4):L041502, 2022.

\bibitem{Sotani:2017pfj}
Hajime Sotani and Kostas~D. Kokkotas.
\newblock {Maximum mass limit of neutron stars in scalar-tensor gravity}.
\newblock {\em Phys. Rev. D}, 95(4):044032, 2017.

\bibitem{kam81}
Burkhard Kampfer.
\newblock {On the Possibility of Stable Quark and Pion Condensed Stars}.
\newblock {\em J. Phys. A}, 14:L471--L475, 1981.

\bibitem{Glendenning:1998ag}
Norman~K. Glendenning and Christiane Kettner.
\newblock {Nonidentical neutron star twins}.
\newblock {\em Astron. Astrophys.}, 353:L9, 2000.

\bibitem{sch00}
K.~Schertler, C.~Greiner, J.~Schaffner-Bielich, and M.~H. Thoma.
\newblock {Quark phases in neutron stars and a 'third family' of compact stars
  as a signature for phase transitions}.
\newblock {\em Nucl. Phys. A}, 677:463--490, 2000.

\bibitem{shaf02}
Jurgen Schaffner-Bielich, Matthias Hanauske, Horst Stoecker, and Walter
  Greiner.
\newblock {Phase transition to hyperon matter in neutron stars}.
\newblock {\em Phys. Rev. Lett.}, 89:171101, 2002.

\bibitem{Novak:1999jg}
Jerome Novak and Jose~M. Ibanez.
\newblock {Gravitational waves from the collapse and bounce of a stellar core
  in tensor scalar gravity}.
\newblock {\em Astrophys. J.}, 533:392--405, 2000.

\bibitem{Novak:1997hw}
Jerome Novak.
\newblock {Spherical neutron star collapse in tensor - scalar theory of
  gravity}.
\newblock {\em Phys. Rev. D}, 57:4789--4801, 1998.

\bibitem{rot20}
Roxana Rosca-Mead, Ulrich Sperhake, Christopher~J. Moore, Michalis Agathos,
  Davide Gerosa, and Christian~D. Ott.
\newblock {Core collapse in massive scalar-tensor gravity}.
\newblock {\em Phys. Rev. D}, 102(4):044010, 2020.

\bibitem{Mendes21}
Raissa F.~P. Mendes, N\'estor Ortiz, and Nikolaos Stergioulas.
\newblock {Nonlinear dynamics of oscillating neutron stars in scalar-tensor
  gravity}.
\newblock {\em Phys. Rev. D}, 104(10):104036, 2021.

\bibitem{chat20}
Katerina Chatziioannou and Sophia Han.
\newblock {Studying strong phase transitions in neutron stars with
  gravitational waves}.
\newblock {\em Phys. Rev. D}, 101(4):044019, 2020.

\bibitem{zhang01}
Bing Zhang and Peter Meszaros.
\newblock {Gamma-ray burst afterglow with continuous energy injection:
  Signature of a highly magnetized millisecond pulsar}.
\newblock {\em Astrophys. J. Lett.}, 552:L35--L38, 2001.

\bibitem{bern20}
Sebastiano Bernuzzi.
\newblock {Neutron Star Merger Remnants}.
\newblock {\em Gen. Rel. Grav.}, 52(11):108, 2020.

\bibitem{sar20}
Nikhil Sarin, Paul~D. Lasky, and Gregory Ashton.
\newblock {Gravitational waves or deconfined quarks: what causes the premature
  collapse of neutron stars born in short gamma-ray bursts?}
\newblock {\em Phys. Rev. D}, 101(6):063021, 2020.

\bibitem{Most:2018eaw}
Elias~R. Most, L.~Jens Papenfort, Veronica Dexheimer, Matthias Hanauske, Stefan
  Schramm, Horst St\"ocker, and Luciano Rezzolla.
\newblock {Signatures of quark-hadron phase transitions in general-relativistic
  neutron-star mergers}.
\newblock {\em Phys. Rev. Lett.}, 122(6):061101, 2019.

\bibitem{Bauswein:2018bma}
Andreas Bauswein, Niels-Uwe~F. Bastian, David~B. Blaschke, Katerina
  Chatziioannou, James~A. Clark, Tobias Fischer, and Micaela Oertel.
\newblock {Identifying a first-order phase transition in neutron star mergers
  through gravitational waves}.
\newblock {\em Phys. Rev. Lett.}, 122(6):061102, 2019.

\bibitem{Weih:2019xvw}
Lukas~R. Weih, Matthias Hanauske, and Luciano Rezzolla.
\newblock {Postmerger Gravitational-Wave Signatures of Phase Transitions in
  Binary Mergers}.
\newblock {\em Phys. Rev. Lett.}, 124(17):171103, 2020.

\bibitem{Liebling:2020dhf}
Steven~L. Liebling, Carlos Palenzuela, and Luis Lehner.
\newblock {Effects of High Density Phase Transitions on Neutron Star Dynamics}.
\newblock {\em Class. Quant. Grav.}, 38(11):115007, 2021.

\bibitem{Most:2019onn}
Elias~R. Most, L.~Jens~Papenfort, Veronica Dexheimer, Matthias Hanauske, Horst
  Stoecker, and Luciano Rezzolla.
\newblock {On the deconfinement phase transition in neutron-star mergers}.
\newblock {\em Eur. Phys. J. A}, 56(2):59, 2020.

\bibitem{Blacker:2020nlq}
Sebastian Blacker, Niels-Uwe~F. Bastian, Andreas Bauswein, David~B. Blaschke,
  Tobias Fischer, Micaela Oertel, Theodoros Soultanis, and Stefan Typel.
\newblock {Constraining the onset density of the hadron-quark phase transition
  with gravitational-wave observations}.
\newblock {\em Phys. Rev. D}, 102(12):123023, 2020.

\bibitem{Berti:2015itd}
Emanuele Berti et~al.
\newblock {Testing General Relativity with Present and Future Astrophysical
  Observations}.
\newblock {\em Class. Quant. Grav.}, 32:243001, 2015.

\bibitem{shao19}
Lijing Shao.
\newblock {Degeneracy in Studying the Supranuclear Equation of State and
  Modified Gravity with Neutron Stars}.
\newblock {\em AIP Conf. Proc.}, 2127(1):020016, 2019.

\bibitem{Sperhake:2017itk}
Ulrich Sperhake, Christopher~J. Moore, Roxana Rosca, Michalis Agathos, Davide
  Gerosa, and Christian~D. Ott.
\newblock {Long-lived inverse chirp signals from core collapse in massive
  scalar-tensor gravity}.
\newblock {\em Phys. Rev. Lett.}, 119(20):201103, 2017.

\bibitem{Gerosa16}
Davide {Gerosa}, Ulrich {Sperhake}, and Christian~D. {Ott}.
\newblock {Numerical simulations of stellar collapse in scalar-tensor theories
  of gravity}.
\newblock {\em Classical and Quantum Gravity}, 33(13):135002, July 2016.

\bibitem{alta17}
Zahra {Altaha Motahar}, Jose~Luis {Bl{\'a}zquez-Salcedo}, Burkhard {Kleihaus},
  and Jutta {Kunz}.
\newblock {Scalarization of neutron stars with realistic equations of state}.
\newblock {\em \prd}, 96(6):064046, September 2017.

\bibitem{Chiba22}
Takeshi {Chiba}.
\newblock {Spontaneous scalarization in scalar-tensor theories with conformal
  symmetry as an attractor}.
\newblock {\em Progress of Theoretical and Experimental Physics},
  2022(1):013E01, January 2022.

\bibitem{Will14}
Clifford~M. Will.
\newblock {The Confrontation between General Relativity and Experiment}.
\newblock {\em Living Rev. Rel.}, 17:4, 2014.

\bibitem{OBoyle20}
Michael~F. O'Boyle, Charalampos Markakis, Nikolaos Stergioulas, and Jocelyn~S.
  Read.
\newblock {Parametrized equation of state for neutron star matter with
  continuous sound speed}.
\newblock {\em Phys. Rev. D}, 102(8):083027, 2020.

\bibitem{LVK}
B.~P. Abbott et~al.
\newblock {GW170817: Measurements of neutron star radii and equation of state}.
\newblock {\em Phys. Rev. Lett.}, 121(16):161101, 2018.

\bibitem{NANOGrav}
H.~T. Cromartie et~al.
\newblock {Relativistic Shapiro delay measurements of an extremely massive
  millisecond pulsar}.
\newblock {\em Nature Astron.}, 4(1):72--76, 2019.

\bibitem{Tuna:2022qqr}
Semih Tuna, K\i{}van\c{c}~\.I. \"Unl\"ut\"urk, and Fethi~M. Ramazano\u{g}lu.
\newblock {Constraining scalar-tensor theories using neutron star mass and
  radius measurements}.
\newblock 4 2022.

\bibitem{Harada98}
Tomohiro Harada.
\newblock {Neutron stars in scalar tensor theories of gravity and catastrophe
  theory}.
\newblock {\em Phys. Rev. D}, 57:4802--4811, 1998.

\bibitem{Mendes:2018qwo}
Raissa F.~P. Mendes and N\'estor Ortiz.
\newblock {New class of quasinormal modes of neutron stars in scalar-tensor
  gravity}.
\newblock {\em Phys. Rev. Lett.}, 120(20):201104, 2018.

\bibitem{Damour92}
Thibault Damour and Gilles Esposito-Farese.
\newblock {Tensor multiscalar theories of gravitation}.
\newblock {\em Class. Quant. Grav.}, 9:2093--2176, 1992.

\bibitem{Horbatsch:2015bua}
Michael Horbatsch, Hector~O. Silva, Davide Gerosa, Paolo Pani, Emanuele Berti,
  Leonardo Gualtieri, and Ulrich Sperhake.
\newblock {Tensor-multi-scalar theories: relativistic stars and 3 + 1
  decomposition}.
\newblock {\em Class. Quant. Grav.}, 32(20):204001, 2015.

\bibitem{Doneva:2020afj}
Daniela~D. Doneva and Stoytcho~S. Yazadjiev.
\newblock {Nontopological spontaneously scalarized neutron stars in
  tensor-multiscalar theories of gravity}.
\newblock {\em Phys. Rev. D}, 101(10):104010, 2020.

\bibitem{piro11}
Anthony~L. {Piro} and Christian~D. {Ott}.
\newblock {Supernova Fallback onto Magnetars and Propeller-powered Supernovae}.
\newblock {\em Astrophys. J.}, 736(2):108, 2011.

\bibitem{ish21}
Wataru Ishizaki, Kunihito Ioka, and Kenta Kiuchi.
\newblock {Fallback Accretion Model for the Years-to-decades X-Ray Counterpart
  to GW170817}.
\newblock {\em Astrophys. J. Lett.}, 916(2):L13, 2021.

\bibitem{van95}
E.~P.~J. {van den Heuvel} and O.~{Bitzaraki}.
\newblock {The magnetic field strength versus orbital period relation for
  binary radio pulsars with low-mass companions: evidence for neutron-star
  formation by accretion-induced collapse?}
\newblock {\em Astron. Astrophys.}, 297:L41, May 1995.

\bibitem{des19}
Dhruv Desai, Brian~D. Metzger, and Francois Foucart.
\newblock {Imprints of r-process heating on fall-back accretion: distinguishing
  black hole\textendash{}neutron star from double neutron star mergers}.
\newblock {\em Mon. Not. Roy. Astron. Soc.}, 485(3):4404--4412, 2019.

\bibitem{lamb73}
F.~K. {Lamb}, C.~J. {Pethick}, and D.~{Pines}.
\newblock {A Model for Compact X-Ray Sources: Accretion by Rotating Magnetic
  Stars}.
\newblock {\em \apj}, 184:271--290, August 1973.

\bibitem{Fujibayashi17}
Sho Fujibayashi, Kenta Kiuchi, Nobuya Nishimura, Yuichiro Sekiguchi, and Masaru
  Shibata.
\newblock {Mass Ejection from the Remnant of a Binary Neutron Star Merger:
  Viscous-Radiation Hydrodynamics Study}.
\newblock {\em Astrophys. J.}, 860(1):64, 2018.

\bibitem{suv21}
A.~G. Suvorov.
\newblock {Ultra-compact X-ray binaries as dual-line gravitational-wave
  sources}.
\newblock {\em Mon. Not. Roy. Astron. Soc.}, 503(4):5495--5503, 2021.

\bibitem{szcz21}
Marek Szczepanczyk et~al.
\newblock {Detecting and reconstructing gravitational waves from the next
  galactic core-collapse supernova in the advanced detector era}.
\newblock {\em Phys. Rev. D}, 104(10):102002, 2021.

\bibitem{prak06}
Madappa Prakash, Ignazio Bombaci, Manju Prakash, Paul~J. Ellis, James~M.
  Lattimer, and Roland Knorren.
\newblock {Composition and structure of protoneutron stars}.
\newblock {\em Phys. Rept.}, 280:1--77, 1997.

\bibitem{glen97}
Norman~K. Glendenning, S.~Pei, and F.~Weber.
\newblock {Signal of quark deconfinement in the timing structure of pulsar
  spindown}.
\newblock {\em Phys. Rev. Lett.}, 79:1603--1606, 1997.

\bibitem{sg21}
Arthur~G. Suvorov and Kostas Glampedakis.
\newblock {Magnetically supramassive neutron stars}.
\newblock {\em Phys. Rev. D}, 105(6):L061302, 2022.

\bibitem{Shibata:2013pra}
Masaru Shibata, Keisuke Taniguchi, Hirotada Okawa, and Alessandra Buonanno.
\newblock {Coalescence of binary neutron stars in a scalar-tensor theory of
  gravity}.
\newblock {\em Phys. Rev. D}, 89(8):084005, 2014.

\bibitem{Christian:2020xwz}
Jan-Erik Christian and J\"urgen Schaffner-Bielich.
\newblock {Supermassive Neutron Stars Rule Out Twin Stars}.
\newblock {\em Phys. Rev. D}, 103(6):063042, 2021.

\bibitem{Deloudis:2021agp}
Themistoklis Deloudis, Polychronis Koliogiannis, and Charalampos Moustakidis.
\newblock {Twin stars: probe of phase transition from hadronic to quark
  matter}.
\newblock {\em EPJ Web Conf.}, 252:06001, 2021.

\bibitem{Espino:2021adh}
Pedro~L. Espino and Vasileios Paschalidis.
\newblock {Fate of twin stars on the unstable branch: Implications for the
  formation of twin stars}.
\newblock {\em Phys. Rev. D}, 105(4):043014, 2022.

\bibitem{Bauswein:2022vtq}
Andreas Bauswein, David Blaschke, and Tobias Fischer.
\newblock {\em {Effects of a strong phase transition on supernova explosions,
  compact stars and their mergers}}.
\newblock 3 2022.

\bibitem{bar13}
Enrico Barausse, Carlos Palenzuela, Marcelo Ponce, and Luis Lehner.
\newblock {Neutron-star mergers in scalar-tensor theories of gravity}.
\newblock {\em Phys. Rev. D}, 87:081506, 2013.

\bibitem{pon14}
Marcelo Ponce, Carlos Palenzuela, Enrico Barausse, and Luis Lehner.
\newblock {Electromagnetic outflows in a class of scalar-tensor theories:
  Binary neutron star coalescence}.
\newblock {\em Phys. Rev. D}, 91(8):084038, 2015.

\bibitem{Sagunski:2017nzb}
Laura Sagunski, Jun Zhang, Matthew~C. Johnson, Luis Lehner, Mairi
  Sakellariadou, Steven~L. Liebling, Carlos Palenzuela, and David Neilsen.
\newblock {Neutron star mergers as a probe of modifications of general
  relativity with finite-range scalar forces}.
\newblock {\em Phys. Rev. D}, 97(6):064016, 2018.

\bibitem{huang19}
Junwu {Huang}, Matthew~C. {Johnson}, Laura {Sagunski}, Mairi {Sakellariadou},
  and Jun {Zhang}.
\newblock {Prospects for axion searches with Advanced LIGO through binary
  mergers}.
\newblock {\em \prd}, 99(6):063013, March 2019.

\bibitem{huang21}
Jun {Zhang}, Zhenwei {Lyu}, Junwu {Huang}, Matthew~C. {Johnson}, Laura
  {Sagunski}, Mairi {Sakellariadou}, and Huan {Yang}.
\newblock {First Constraints on Nuclear Coupling of Axionlike Particles from
  the Binary Neutron Star Gravitational Wave Event GW170817}.
\newblock {\em \prl}, 127(16):161101, October 2021.

\bibitem{oko20}
Maria Okounkova.
\newblock {Numerical relativity simulation of GW150914 in Einstein dilaton
  Gauss-Bonnet gravity}.
\newblock {\em Phys. Rev. D}, 102(8):084046, 2020.

\bibitem{oko20b}
Maria Okounkova.
\newblock {Numerical relativity simulation of GW150914 in Einstein dilaton
  Gauss-Bonnet gravity}.
\newblock {\em Phys. Rev. D}, 102(8):084046, 2020.

\bibitem{abb17}
B.~P. Abbott et~al.
\newblock {GW170817: Observation of Gravitational Waves from a Binary Neutron
  Star Inspiral}.
\newblock {\em Phys. Rev. Lett.}, 119(16):161101, 2017.

\bibitem{hou18}
Shu-Jin Hou, Tong Liu, Ren-Xin Xu, Hui-Jun Mu, Cui-Ying Song, Da-Bin Lin, and
  Wei-Min Gu.
\newblock The x-ray light curve in grb 170714a: Evidence for a quark star?
\newblock {\em Astrophys. J.}, 854(2):104, 2018.

\bibitem{gal98}
T.~J. {Galama}, P.~M. {Vreeswijk}, J.~{van Paradijs}, C.~{Kouveliotou},
  T.~{Augusteijn}, H.~{B{\"o}hnhardt}, J.~P. {Brewer}, V.~{Doublier}, J.~F.
  {Gonzalez}, B.~{Leibundgut}, C.~{Lidman}, O.~R. {Hainaut}, F.~{Patat},
  J.~{Heise}, J.~{in't Zand}, K.~{Hurley}, P.~J. {Groot}, R.~G. {Strom}, P.~A.
  {Mazzali}, K.~{Iwamoto}, K.~{Nomoto}, H.~{Umeda}, T.~{Nakamura}, T.~R.
  {Young}, T.~{Suzuki}, T.~{Shigeyama}, T.~{Koshut}, M.~{Kippen},
  C.~{Robinson}, P.~{de Wildt}, R.~A.~M.~J. {Wijers}, N.~{Tanvir},
  J.~{Greiner}, E.~{Pian}, E.~{Palazzi}, F.~{Frontera}, N.~{Masetti},
  L.~{Nicastro}, M.~{Feroci}, E.~{Costa}, L.~{Piro}, B.~A. {Peterson},
  C.~{Tinney}, B.~{Boyle}, R.~{Cannon}, R.~{Stathakis}, E.~{Sadler}, M.~C.
  {Begam}, and P.~{Ianna}.
\newblock {An unusual supernova in the error box of the
  {\ensuremath{\gamma}}-ray burst of 25 April 1998}.
\newblock {\em \nat}, 395(6703):670--672, October 1998.

\bibitem{Reitze:2019iox}
David Reitze et~al.
\newblock {Cosmic Explorer: The U.S. Contribution to Gravitational-Wave
  Astronomy beyond LIGO}.
\newblock {\em Bull. Am. Astron. Soc.}, 51(7):035, 2019.

\bibitem{Punturo:2010zz}
M.~Punturo et~al.
\newblock {The Einstein Telescope: A third-generation gravitational wave
  observatory}.
\newblock {\em Class. Quant. Grav.}, 27:194002, 2010.

\bibitem{Hild:2010id}
S.~Hild et~al.
\newblock {Sensitivity Studies for Third-Generation Gravitational Wave
  Observatories}.
\newblock {\em Class. Quant. Grav.}, 28:094013, 2011.

\bibitem{Maggiore:2019uih}
Michele Maggiore et~al.
\newblock {Science Case for the Einstein Telescope}.
\newblock {\em JCAP}, 03:050, 2020.

\bibitem{sag09}
I.~Sagert, T.~Fischer, M.~Hempel, G.~Pagliara, J.~Schaffner-Bielich,
  A.~Mezzacappa, F.~K. Thielemann, and M.~Liebendorfer.
\newblock {Signals of the QCD phase transition in core-collapse supernovae}.
\newblock {\em Phys. Rev. Lett.}, 102:081101, 2009.

\bibitem{qin98}
Bo~Qin, Xiang-Ping Wu, Ming-Chung Chu, and Li-Zhi Fang.
\newblock {The collapse of neutron stars in high mass binaries as the energy
  source for the gamma-ray bursts}.
\newblock {\em Astrophys. J. Lett.}, 494:L57, 1998.

\bibitem{perna21}
Rosalba Perna, Hiromichi Tagawa, Zoltan Haiman, and Imre Bartos.
\newblock {Accretion-Induced Collapse of Neutron Stars in the Disks of Active
  Galactic Nuclei}.
\newblock {\em Astrophys. J.}, 915(1):10, 2021.

\bibitem{GR1D}
Evan {O'Connor} and Christian~D. {Ott}.
\newblock {A new open-source code for spherically symmetric stellar collapse to
  neutron stars and black holes}.
\newblock {\em Classical and Quantum Gravity}, 27(11):114103, June 2010.

\bibitem{Mendes16}
Raissa F.~P. {Mendes} and N{\'e}stor {Ortiz}.
\newblock {Highly compact neutron stars in scalar-tensor theories of gravity:
  Spontaneous scalarization versus gravitational collapse}.
\newblock {\em \prd}, 93(12):124035, June 2016.

\bibitem{Cheong19}
Patrick Chi-Kit {Cheong} and Tjonnie Guang~Feng {Li}.
\newblock {Numerical studies on core collapse supernova in self-interacting
  massive scalar-tensor gravity}.
\newblock {\em \prd}, 100(2):024027, July 2019.

\bibitem{gill72}
P.~E. Gill and G.~F. Miller.
\newblock {An Algorithm for the Integration of Unequally Spaced Data}.
\newblock {\em The Computer Journal}, 15(1):80--83, 1972.

\bibitem{East21}
William~E. {East} and Justin~L. {Ripley}.
\newblock {Evolution of Einstein-scalar-Gauss-Bonnet gravity using a modified
  harmonic formulation}.
\newblock {\em \prd}, 103(4):044040, February 2021.

\bibitem{dimmel09}
Harald {Dimmelmeier}, Michal {Bejger}, Pawel {Haensel}, and J.~Leszek {Zdunik}.
\newblock {Dynamic migration of rotating neutron stars due to a phase
  transition instability}.
\newblock {\em \mnras}, 396(4):2269--2288, July 2009.

\bibitem{Falcone21}
Riccardo {Falcone}, Daniela~D. {Doneva}, Kostas~D. {Kokkotas}, and Stoytcho~S.
  {Yazadjiev}.
\newblock {Nonlinear stability of soliton solutions for massive
  tensor-multiscalar theories}.
\newblock {\em \prd}, 104(6):064045, September 2021.

\bibitem{kuan21}
Hao-Jui {Kuan}, Jasbir {Singh}, Daniela~D. {Doneva}, Stoytcho~S. {Yazadjiev},
  and Kostas~D. {Kokkotas}.
\newblock {Nonlinear evolution and nonuniqueness of scalarized neutron stars}.
\newblock {\em \prd}, 104(12):124013, December 2021.

\bibitem{Rosswog22}
Stephan {Rosswog}.
\newblock {Modelling astrophysical fluids with particles}.
\newblock {\em arXiv e-prints}, page arXiv:2201.05896, January 2022.

\bibitem{Rosca-Mead:2020bzt}
Roxana Rosca-Mead, Christopher~J. Moore, Ulrich Sperhake, Michalis Agathos, and
  Davide Gerosa.
\newblock {Structure of neutron stars in massive scalar-tensor gravity}.
\newblock {\em Symmetry}, 12(9):1384, 2020.

\end{thebibliography}

\appendix
\section{Reduced field equations}\label{appendix.A}

\begin{table*}
	\centering
	\caption{Coefficients of the ansatz catered to EOS APR4 (cf.~Tabs.~II and III of Appendix B in \cite{OBoyle20}).}
	\begin{tabular}{lcccc}
	    \hline
	    \hline
		$\rho_i$ (g/cm$^{3}$) & $\Gamma_i$ & $K_i$ (cgs) & $\Lambda_i$ (g/cm$^{3}$) & $a_i$ \\
		\hline
		0 & 1.611 & $5.214\times10^{-9}$ & 0 & 0 \\
	    \hline
	    $6.285\times10^5$ & 1.440 & $5.726\times10^{-8}$ & -1.354 & $-1.861\times10^{-5}$\\
	    \hline 
	    $1.826\times10^8$ & 1.269 & $1.662\times10^{-6}$ & $-6.025\times10^3$ & $-5.278\times10^{-4}$\\
	    \hline 
	    $3.350\times10^{11}$ & -1.841 & $-7.957\times10^{29}$ & $1.193\times10^9$ & $1.035\times10^{-2}$\\
	    \hline 
	    $5.317\times10^{11}$ & 1.382 & $1.746\times10^{-8}$ & $7.077\times10^8$ & $8.208\times10^{-3}$\\
	    \hline 
	    $1.096\times10^{14}$ & 3.169 & $6.166\times10^{-34}$ & $2.499\times10^{11}$ & $2.022\times10^{-2}$\\
	    \hline 
	    $7.413\times10^{14}$ & 3.452 & $3.505\times10^{-38}$ & $6.960\times10^{12}$ & $3.884\times10^{-2}$\\
	    \hline 
	    $9.772\times10^{14}$ & 3.310 & $4.914\times10^{-36}$ & $-1.407\times10^{12}$ & $2.157\times10^{-2}$\\
	    \hline
	\end{tabular}
	\label{tab:eos}
\end{table*}

For the spherically-symmetric Jordan frame metric ansatz adopted here, viz.
\begin{align}
	\tilde{g}_{\mu\nu}=\text{diag} [-\alpha^{2},X^{2},A(\varphi)^2r^{2},A(\varphi)^2r^{2}\sin^{2}\theta ],
\end{align}
the conformally-related, Einstein-frame metric reads
\begin{align}\label{eq:EinMetric}
	g_{\mu\nu}&=\text{diag} [-A(\varphi)^{-2}\alpha^{2},A(\varphi)^{-2}X^{2},r^{2},r^{2}\sin^{2}\theta ] \nonumber\\
	&:=\text{diag}[-e^{2\Phi},e^{2\Lambda},r^{2},r^{2}\sin^{2}\theta],
\end{align}
where the local mass $m(r)$ is related to the metric potential $\Lambda$ via
\begin{align}
	e^{2\Lambda} = \bigg(1-\frac{2m}{r}\bigg)^{-1}.
\end{align}
The form \eqref{eq:EinMetric} reflects our choice of the radial gauge together with polar slicing condition (see, e.g., the original paper reporting the code \texttt{GR1D}\cite{GR1D}).
The corresponding `Einstein' equations then read (see also Eqs.~(2.21)-(2.23) in \cite{Gerosa16} together with Eq.~(6) in \cite{Sperhake:2017itk})
\begin{subequations}
\begin{align}
	&m' = 4\pi r^{2}A(\varphi)^4\bigg(\frac{\rho h}{1-v^2}-P\bigg)+r^{2}\bigg[\frac{A(\varphi)^2}{2}(\psi^{2}+\eta^{2})+V\bigg], \label{eq:m_prm}\\
	&\dot{m}=r^{2}\frac{\alpha}{X}\bigg[ A(\varphi)^2\psi\eta-4\pi A(\varphi)^4 \bigg(\frac{\rho hv}{1-v^2}\bigg) \bigg], \label{eq:m_dot}\\
\intertext{and}
	&\Phi'=X^{2}\bigg[\bigg( \frac{m}{r^{2}}-rV\bigg)A(\varphi)^{-2}
	+4\pi r^{2}A(\varphi)^2\bigg(\frac{\rho hv^2}{1-v^2}+P\bigg)\nonumber\\
	&\qquad+\frac{r}{2}(\psi^{2}+\eta^{2}) \bigg],
	\label{eq:phi_prm}
\end{align}
\end{subequations}
for density $\rho$, pressure $P$, enthalpy $h$, matter's radial speed $v$, and 4-velocity
\begin{align}
	u^{\mu} = \frac{1}{\sqrt{1-v^2}}(\alpha^{-1},vX^{-1},0,0).
\end{align}
We note that the hydrodynamical quantities are evaluated in the Jordan frame, while the metric functions $m$ and $\Phi$ are defined in the Einstein one. In addition, we have introduced two auxiliary variables $\psi$ and $\eta$, respectively, defined as 
\begin{align}
	\psi = \frac{1}{\alpha}\frac{\partial\varphi}{\partial t},\quad\text{and}\quad
	\eta = \frac{1}{X}\frac{\partial\varphi}{\partial r}.
\end{align}

\begin{figure}
	\centering\hspace*{-3mm}
	\includegraphics[width=\columnwidth]{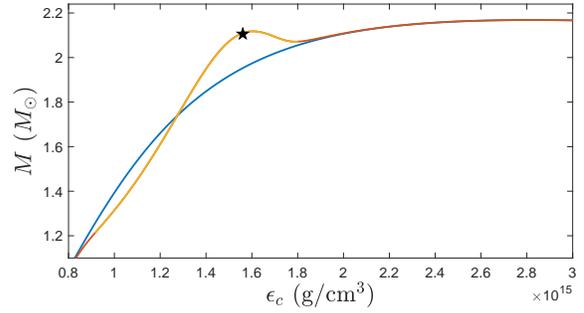}
	\caption{Hydrostatic, APR4 equilibria for $\alpha_0=10^{-2}$, $\beta_0=-5$ and $m_\varphi=10^{-14}$~eV. The scalarized branch is divided in two parts, with the orange one being strongly-scalarized and the red one possessing a very weak scalar field, which virtually coincides with the GR case ($\alpha=0$; blue curve). The black star represents the initial state for the simulations described here.}
	\label{fig:eqrb}
\end{figure}

The enthalpy comprises energy density $\epsilon$ and pressure, given by
\begin{align}
	h=\epsilon+P.
\end{align}
For cold EOS, $\epsilon$ and P are functions of $\rho$; in particular, the piecewise polytropic approximation of EOS, which is adopted in this Letter for the APR4 EOS (see below), gives the relations $\epsilon(\rho)$ and $P(\rho)$ as Eqs.~(4.4) and (4.5) of \cite{OBoyle20}.
The gravitational mass of the star is technically defined as the limit of $m$ as $r$ goes to infinity, though we approximate it by using the value at the outermost grid at $r=9\times10^5$ km; the difference from this mass estimate to that obtained using radii ten times closer to the center of the star, $r=9\times10^4$ km, is only $10^{-5}M_{\odot}$. The neutron star equilibrium solutions for this specific EOS and for certain representative values of the STT parameters employed in the main text of the paper, are plotted in Fig.~\ref{fig:eqrb}. GR solutions are presented as well (blue curve) for comparison. The initial state of our numerical simulation is shown by the black marker at $M = 2.11M_{\odot} $ with a central energy density of $\epsilon_{c} = 1.56 $ g/cm$^{3}$.

The equation for $\varphi$ and the hydrodynamic equations, respectively, given by
\begin{align}\label{eq:KG}
	\square\varphi &= -4\pi \frac{d\ln A}{d\varphi}T+\frac{dV}{d\varphi}
	=-4\pi(\alpha_0+\beta_0\varphi)T+m_{\varphi}^2\varphi,
\end{align}
and 
\begin{align} \label{eq:euler}
	0 = \nabla_{\mu}T^{\mu\nu}-\frac{d\ln A}{d\varphi} T\nabla^{\nu}\varphi,
\end{align}
can be decomposed to a set of first-order hyperbolic differential equations in a conservative-flux form for the variables $\eta$, $\psi$, and the matter quantities 
\begin{align}
    D= \frac{\rho e^{\Lambda}}{\sqrt{1-v^2}}, \,\, S^r= \frac{\rho h v}{1-v^2}, \,\, \tau= \frac{\rho h }{1-v^2} - p -D.
\end{align}
The equations are long and not directly relevant to our end goals, so we refer the reader to the Section II.A of \cite{Mendes16} for the explicit expressions.

\section{Piecewise-polytropic approximation of equation of state}\label{appendix.B}
In this Letter, we adopt the modified piecewise-polytropic approximation for tabulated EOS APR4, where the construction of the approximation is designed such that the sound speed is continuous throughout the star \cite{OBoyle20}. By fusing eight segments at seven benchmark baryonic mass densities, denoted by $\rho_i$, the ansatz,
\begin{align}
	&\epsilon(\rho) =\frac{K_i}{\Gamma_i-1}\rho^{\Gamma_i}+(1+a_i)\rho-\Lambda_i,\\
	&p(\rho) = K_i\rho^{\Gamma_i}+\Lambda_i,
\end{align}
is used for $\rho\in[\rho_{i-1},\rho_i]$ (i.e., the $i$-th segment) to fit the tabulated EOS with parameters $\Gamma_i$. Other parameters ($K_i$, $a_i$, and $\Lambda_i$) are determined by demanding that (i) $p(\rho)$ is differentiable everywhere (fixing $K_{i}$ and $\Lambda_i$), and (ii)  $\epsilon(\rho)$ is differentiable (fixing $a_i$) at the dividing density between the $i$ and $i+1$ segments. The numerical values of the fitting coefficients for APR4 can be found in \cite{OBoyle20} and are listed for convenience in Tab.~\ref{tab:eos}.

\section{Numerical Detail}\label{appendix.C}
The numerical code used in this Letter is modified from the \texttt{GR1D} code \cite{GR1D}, and is thus a variant of the one used in \cite{Cheong19}.
In the code, we use a 3rd order Runge-Kutta method with Courant–Friedrichs–Lewy (CFL) factor of 0.25 to advance these fields between time steps, $t_{k}$ and $t_{k+1}$. The metric functions are updated at the new time step via
\begin{align}
    m(r,t_{k+1}) = \int_{0}^{r} dr' \frac {\partial m(r',t_{k})} {\partial r'},
\end{align}
and
\begin{align}
    \Phi(r,t_{k+1}) = \int_{0}^{r} dr' \frac {\partial \Phi(r',t_{k})} {\partial r'},
\end{align}
where the integration is executed by the NAG package D01GAF that is based on Gill \& Miller's algorithm \cite{gill72}.
Energy is always conserved, in the sense that the energy radiated away by the scalar field,
\begin{align}
    E^{\text{(scalar)}}_{\text{GW}} = \int \mathcal{L}_{\varphi} dt,
\end{align}
and the (gravitational) mass of the star throughout the simulation maintain a nearly constant value, $M-E^{\text{(scalar)}}_{\text{GW}}$, with an error of $\lesssim 10^{-6}M_{\odot}$. Here the flux of scalar radiation is quantified by (e.g., Eq.~(20) of \cite{East21})
\begin{align}
    \mathcal{L}_{\text{GW}}^{\text{(scalar)}}(r)=&\int\sqrt{-\tilde{g}} \left(T_{tr}^{(\varphi)}\partial^{t}\partial^{r}\right)r^2d\Omega\nonumber\\
    =& X\alpha\psi\eta r^2,
\end{align}
where $T_{tr}^{(\varphi)}$ is the scalar contribution to the stress-energy tensor. We note that energy conservation is monitored in the Einstein frame since both the gravitational mass and the scalar flux are evaluated in this frame.

\begin{figure}
	\centering
	\includegraphics[width=\columnwidth]{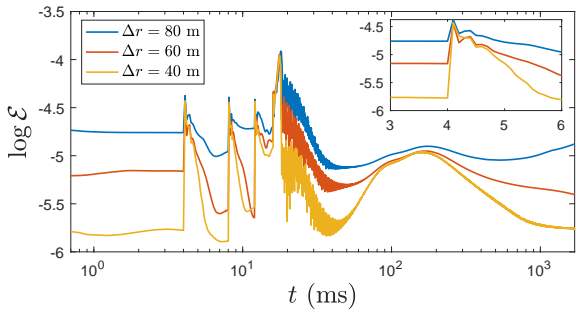}
	\caption{Residual of the constraint equation \eqref{eq:m_dot}, as defined in Eq.~\eqref{eq:error}, as functions of time for several resolutions. Type I accretion  is considered for concreteness.}
	\label{fig:convergence}
\end{figure}

In addition, the redundant equation \eqref{eq:m_dot} can serve as a constraint for the solution at each step, which we use to quantify the convergence of our numerical results; in particular, we define $\mathcal{E}$ to be the $L^2$ norm of the difference of both sides of Eq.~\eqref{eq:m_dot}, viz.
\begin{align}\label{eq:error}
    \mathcal{E}^2 = \int dr \left| \dot{m}-r^{2}\frac{\alpha}{X}\left[ A(\varphi)^2\psi\eta-4\pi A(\varphi)^4 \left(\frac{\rho hv}{1-v^2}\right) \right] \right|^2,
\end{align}
and show in Fig.~\ref{fig:convergence} the evolution of Eq.~\eqref{eq:error} at three resolutions with grid width being 80, 60, and 40 meters, respectively. The logarithmic value of $\mathcal{E}$ decreases by more than 1 when the resolution ($\Delta r=80$~m) is doubled ($\Delta r=40$~m) except for a short period after receiving one clump of inflow (see below), implying that convergence is better than second order. We note that the convergence of the original code is also approximately of second order, as reported in \cite{Gerosa16}.

We simulate the accretion by artificially superposing bulks on the background matter. In particular, the bulk is assumed to have a Gaussian distribution that can be characterised by three parameters, namely the amplitude, mean position, and one sigma width. We recall that the profile we use in the main text has a mean position of $r=0.9 R_{\star}$ for the stellar radius $R_{\star}$, a width of $1$ km, and with amplitude chosen so that the mass of one bulk is $\sim0.004M_{\odot}$. In addition, as will be discussed in detail below, we consider another type of accretion, where the mean position is at $r=1.2R_{\star}$, width is $1$ km, and the amplitude is set to the value such that one bulk weights $\sim0.004M_{\odot}$ (i.e., the bulk is instead injected to a region outside of the star).

Although a synthetic accretion introduces a momentary violation of constraint equation \eqref{eq:m_dot}, reflected by the sudden loss of convergence shown as expected (peaks in Fig.~\ref{fig:convergence}), the infringement is small enough -- due to the minute amount of accreted matter -- that the stability of code keeps the run in good repair. In particular, in the inset we magnify the residual of constraint equation near the first bulk at 4 ms, where we see the loss of convergence is fixed at $\sim5$ ms, i.e., within $\sim1$ ms. The convergence order therefore recovers before receiving the next bulk or the completion of descalarization. Superposing material like this has been used to test stability of neutron stars in the literature, e.g., \cite{dimmel09,Falcone21,kuan21,Espino:2021adh,Rosswog22}; here, we adopt this `trick' to mimic accretion.

\section{Results for Type II accretion}\label{appendix.D}

\begin{figure}
	\centering
	\includegraphics[width=\columnwidth]{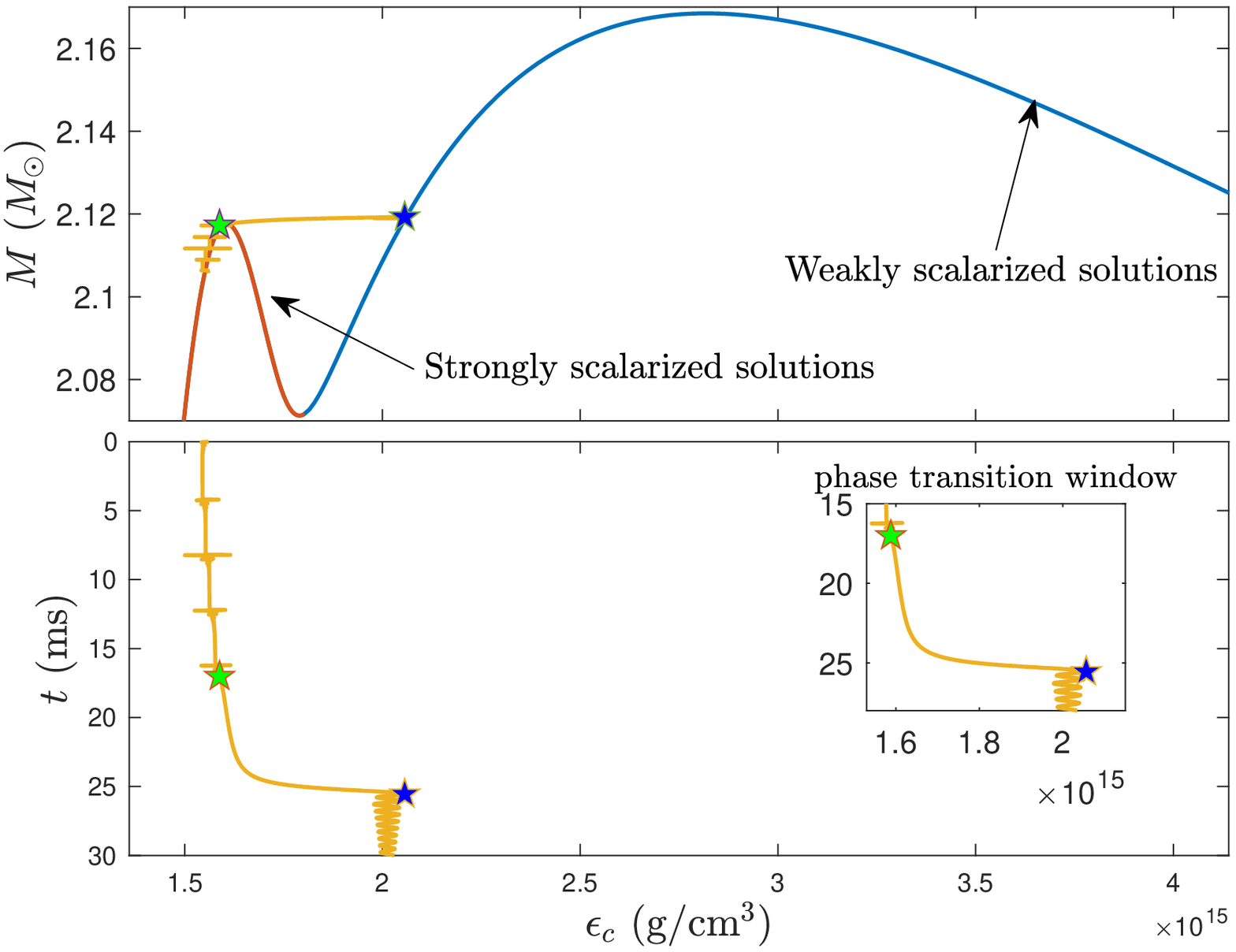}
	\includegraphics[width=\columnwidth]{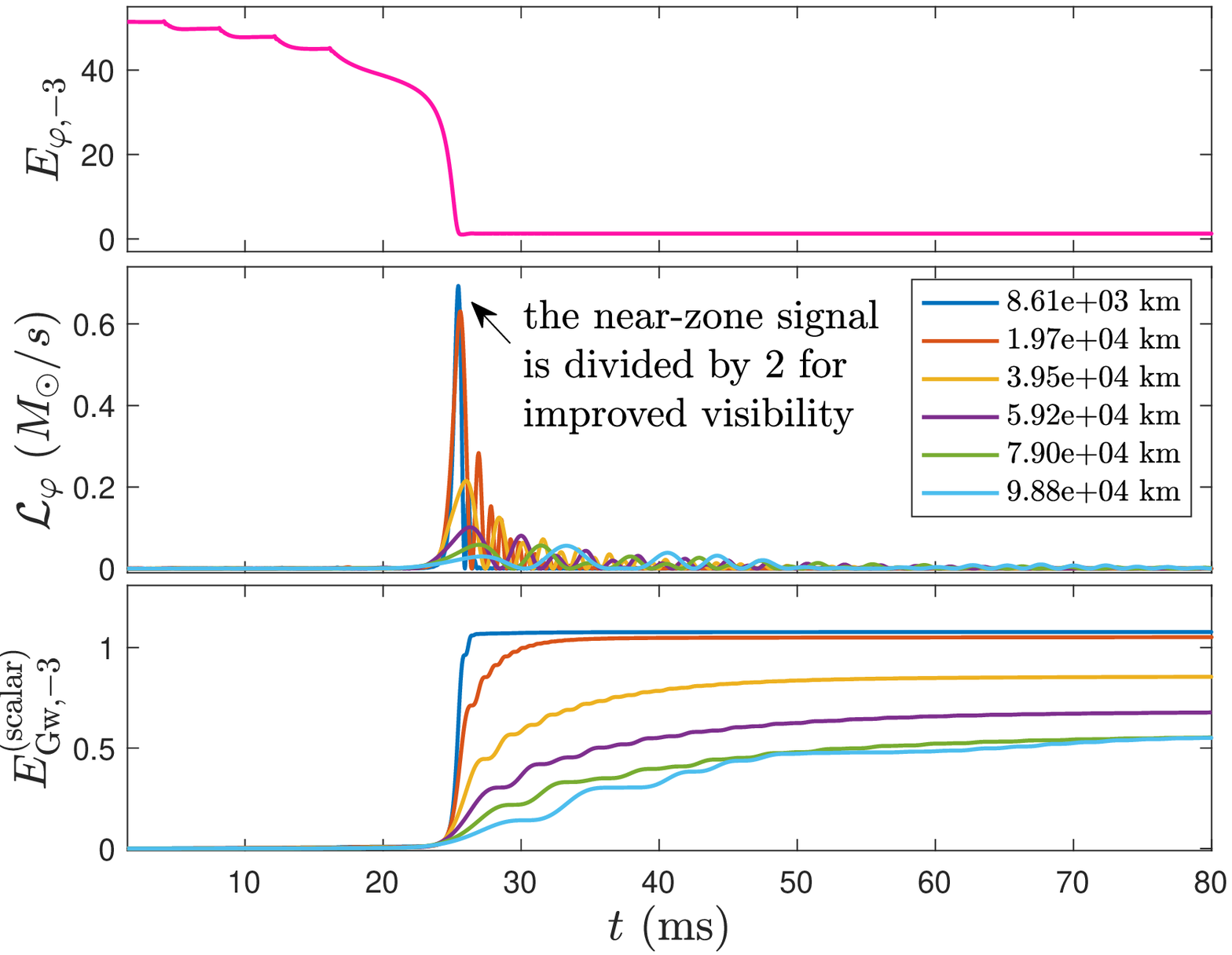}
	\caption{\emph{top panel}: Evolutionary track of a near-critical scalarized star under Type II accretion. \emph{bottom panel}: Evolution of scalar energy under Type II accretion. Here $E_{\varphi,-3}$ and $E^{\text{(scalar)}}_{\varphi,-3}$ are given in the unit of $10^{-3}M_{\odot}$.}
	\label{fig:TypeII1}
\end{figure}
\begin{figure}
	\centering
	\includegraphics[width=\columnwidth]{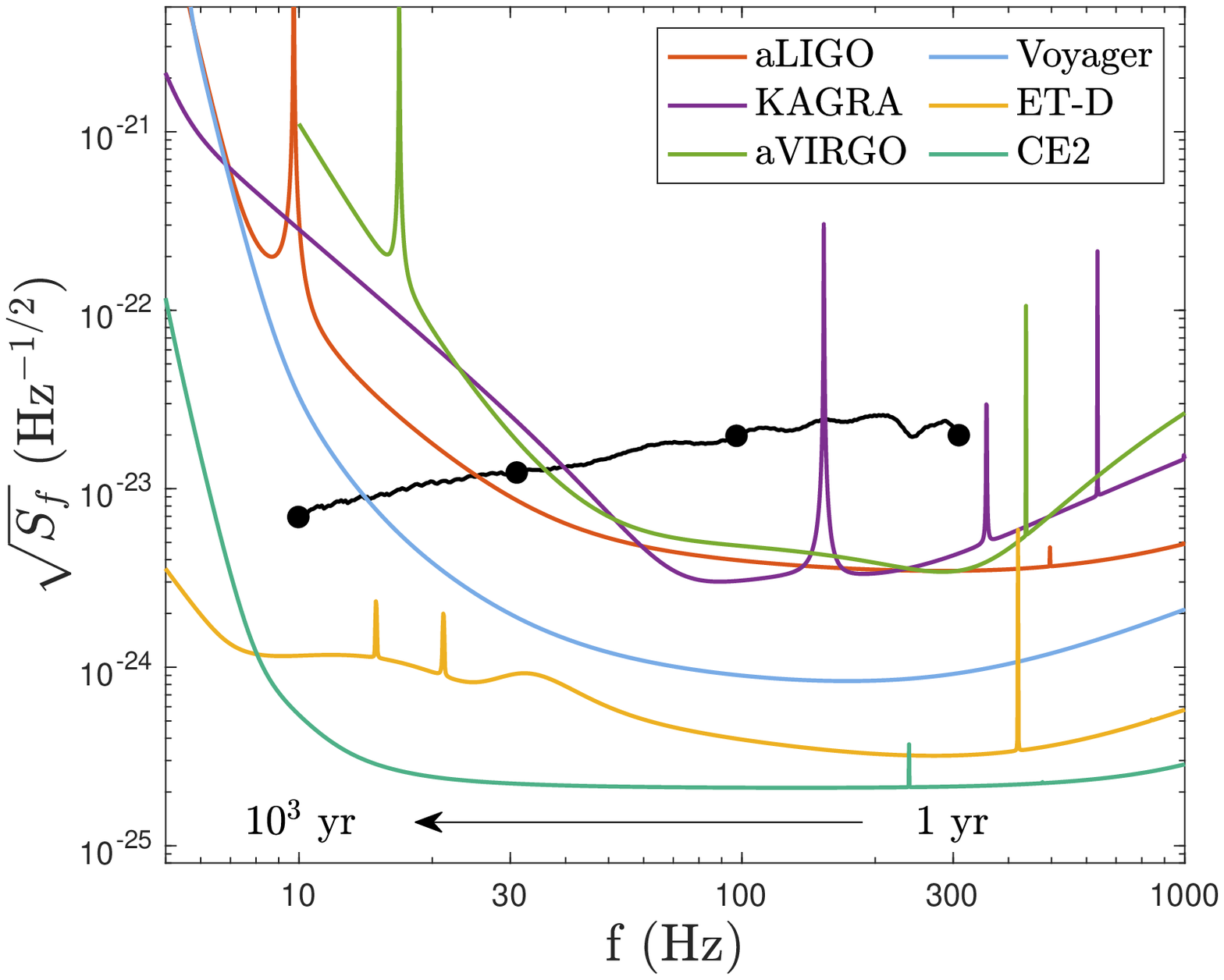}
	\caption{PSD of the scalar-induced GW mode [Eq.~(3) in the main Letter] for Type II accretion, extracted at $L=10$ kpc and a observation time of $T$ = 2 months for retarded time ranging for 1 to $10^3$ years. The $k$-th notch from the right on the curve denotes $10^{(k-1)}$ years.}
	\label{fig:TypeII2}
\end{figure}

\begin{figure}
	\centering
	\includegraphics[width=\columnwidth]{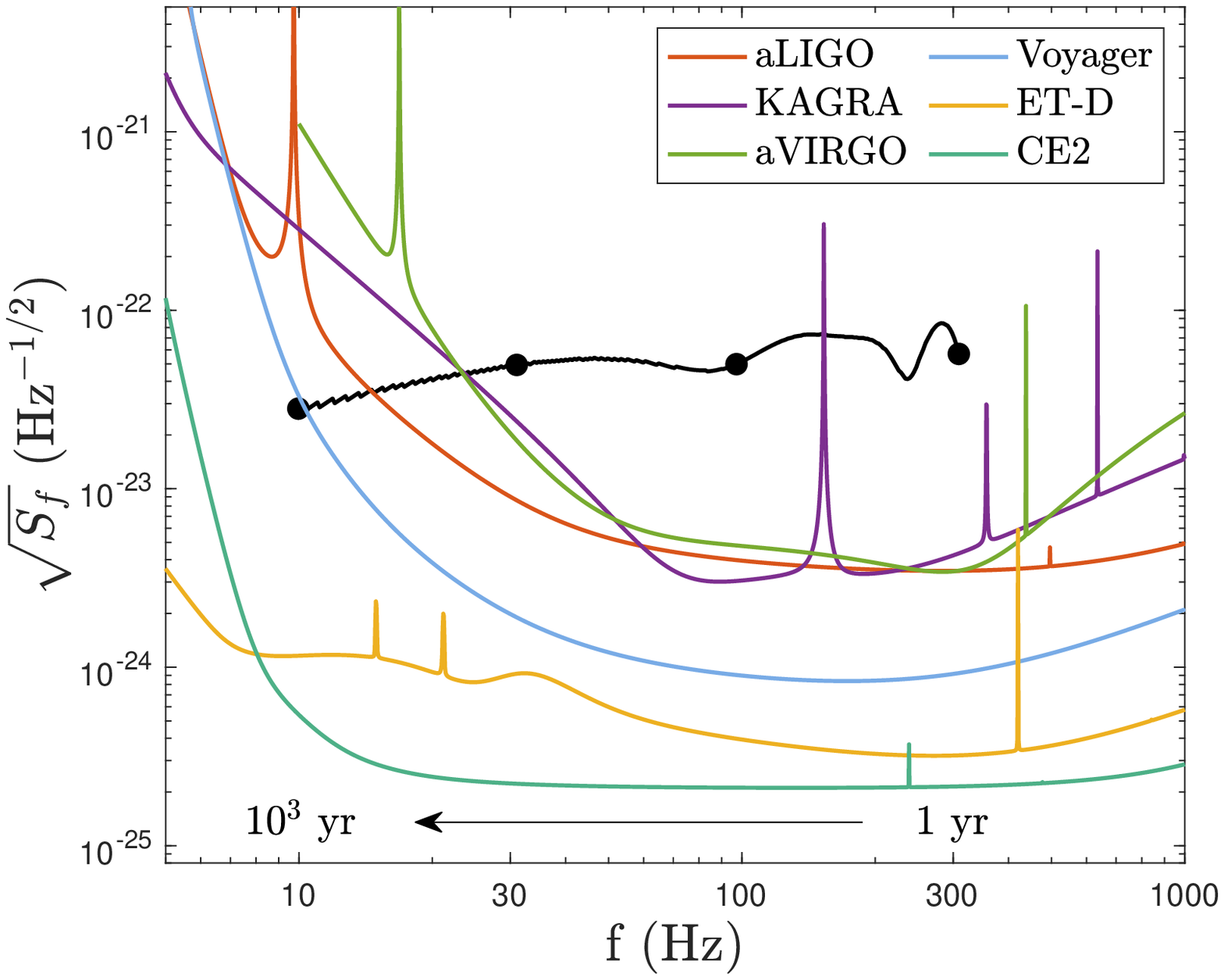}
	\includegraphics[width=\columnwidth]{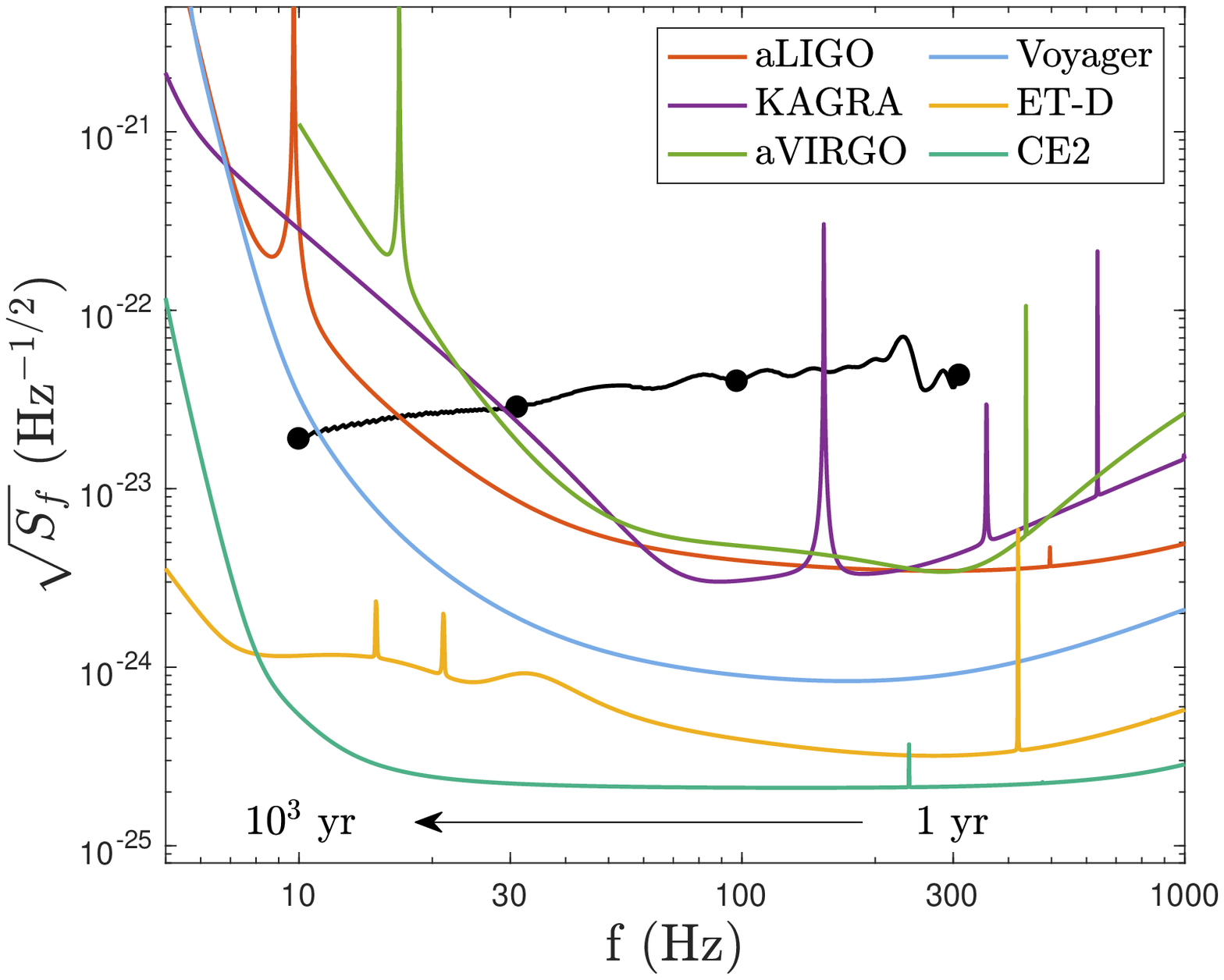}
	\caption{PSD of the scalar-induced GW mode [Eq.~(3) in the main Letter] for second coupling function [Eq.~\eqref{eq:coup2nd}] with Tpye I (top) and Type II (bottom) accretion, extracted at $L=10$ kpc and a observation time of $T$ = 2 months for retarded time ranging for 1 to $10^3$ years. The $k$-th notch from the right on the curve denotes $10^{(k-1)}$ years.}
	\label{fig:coup2nd_strain}
\end{figure}

Here discuss the evolution of a scalarized star undergoing a second type (`Type II') of accretion, introduced above.
The accretion is terminating when a total (baryon) mass of $0.014M_{\odot}$ has been added (after $16.1$ ms). The idea is to demonstrate that the basic conclusions we have made in the paper are generic, and only weakly depend on the particulars of the (short-lived) accretion. We find a similar accretion rate, $\simeq 0.87M_{\odot}\text{s}^{-1}$, as that of the accretion profile used in the main text, but here the descalarization lasts $\sim 8.54$ ms.

In Fig.~\ref{fig:TypeII1}, we plot the evolution track of the star undergoing Type II accretion (top panel), and the scalar emission during the descalarization (bottom panel), while the GW strain caused by the emanating scalar field is presented in Fig.~\ref{fig:TypeII2}. We see that the process is both qualitatively and quantitatively similar to that of the case presented in the main text, which indicates that for the two distinct types of accretion we get practically the same results. This provides some degree of confidence that the process is relatively independent of the initial setup. 

It is natural also to ask whether the change of some theory parameters, namely $\alpha_0$ and $m_\varphi$, will influence significantly our results. The equilibrium with a fixed mass would not change much for sufficiently small $\alpha_0$ (cf.~Fig.~2 of \cite{Rosca-Mead:2020bzt}) and $m_{\varphi}$ (cf.~Fig.~1 of \cite{Ramazanoglu16}); therefore, the scalar-shedding processes for a range of $\alpha_0$ and $m_{\varphi}$ would be similar to each other.
Of course, what scales with $\alpha_0$ is the effective power-spectral density (PSD) of the resulting scalar radiation that is discussed in the main text of the letter. The mass similarly scales the extent to which dispersive stretching occurs.

\section{Another Coupling Function}\label{appendix.E}
To emphasise that our results are general with respect to different coupling functions (at least those having been considered in the literature), we repeat the accretion scenario with the aforementioned types for a particular scalarized star in the theory defined by the coupling function \cite{alta17}
\begin{align}\label{eq:coup2nd}
    A(\varphi) &= \frac{e^{\alpha_0\varphi}}{\cosh({\sqrt{-\beta_0}}\varphi)}.
\end{align}
The Taylor expansion of the logarithmic derivative of the coupling above \eqref{eq:coup2nd} reads
\begin{align}
    \frac{d\ln A(\varphi)}{d\varphi} = \alpha_0+\beta_0\varphi+\frac{\beta_0^2}{3}\varphi^3+O(\varphi^5),
\end{align}
where the first two terms coincides with the coupling function used in the rest of the simulations, that is $\alpha(\varphi)=d\ln A(\varphi)/d\varphi = \alpha_0 + \beta_0 \varphi$. Thus, for a particular choice of $\alpha_0$ and $\beta_0$, these two coupling functions match up to the linear term, and only differ in the cubic contribution. Since the scalar field in our simulations is sufficiently small, the cubic and higher order terms do not play a significant role. This will be indeed demonstrated below.

For the parameters $\alpha_0=10^{-2}$, $\beta_0=-5$, and $m_\varphi=10^{-14}$ eV, we simulate the process that a star with the mass of $2.063$ undergoing type I and II accretions with the coupling \eqref{eq:coup2nd}. 
The resulting strains of scalar-induced GWs are presented in Fig.~\ref{fig:coup2nd_strain}, similar to those shown above and in the main Letter: a gravitational phase transition is realised when the star becomes over-accreted compared to the maximum mass of the scalarized solutions ($2.081M_{\odot}$ for the considered model). This will eventually lead to continuous scalar-radiation that may be observed with strain of $\gtrsim10^{-22} (\text{kpc}/L)^{3/2} \text{ Hz}^{-1/2}$ at frequencies of $\lesssim300 \text{ Hz}$ as shown in Figs.~\ref{fig:TypeII2} and \ref{fig:coup2nd_strain}.

If one compares closely Figs.~\ref{fig:TypeII2} and \ref{fig:coup2nd_strain}, the strain differs by a factor $\sim 2$ one decade after the onset of emissions. Similar differences are seen at later times as well. Despite the small difference in the strength of the signal, which depends also on the initial states (e.g., baryon mass) of both theories, the accretion-induced descalarization occurs, and proves that the results presented in the Letter are not tied to a certain coupling function. Instead, they capture the general picture for a large class of couplings.

\end{document}